\documentstyle[osa,manuscript,tranmacros,fleqn,graphicx]{revtex}

\newcommand{\be}{\begin{equation}}
\newcommand{\ee}{\end{equation}}
\newcommand{\bea}{\begin{eqnarray*}} 
\newcommand{\eea}{\end{eqnarray*}}
\begin{document}                

\title{Excitation of low-n TAE instabilities by energetic particles 
in global gyrokinetic tokamak plasmas}

\author
{
Y.~Nishimura 
\footnote{
Previous address : Department of Physics and Astronomy, University of California, Irvine, California 92697, USA}
}
\address{Plasma and Space Science Center, National Cheng Kung University, Tainan 70101, Taiwan}

\maketitle
\begin{abstract}
The first linear global electromagnetic gyrokinetic particle simulation on the 
excitation of toroidicity induced Alfven eigenmode (TAE) by energetic particles is reported. 
With an increase in the energetic particle pressure, 
the TAE frequency moves down into the lower continuum.

\vspace{0.5cm}

\end{abstract} 

\vspace{0.5cm}

\indent
Toroidicity induced Alfven eigenmode (TAE)\cite{che85,che86,fu89}
can play important roles in burning plasmas.
The TAE modes can be excited when energetic particles, for example
fusion born alpha particles, resonate with the phase velocity
of the shear Alfven wave which resides within the frequency 
gap of the Alfven continuum.

Shear Alfven wave oscillations, 
continuum damping, and the appearance of the frequency gap in
toroidal geometries by gyrokinetic particle simulation 
have been recently reported.\cite{nis07a} 
The simulation of Ref.4 is demonstrated in the long wavelength 
magnetohydrodynamic (MHD) like limit in the absence of kinetic ions.
In this letter, taking exactly the same parameters\cite{fu89,nis07a}
but adding the energetic ion particles, the first 
linear particle simulation on the excitation of the TAE modes is reported.
The simulation is done without employing MHD equation.
The simulation is not the conventional gyrokinetic-MHD hybrid ones,\cite{par92,fu95,che91}
where the kinetic ions enter the system through the stress tensor.
The setting of the simulation is kept as faithful as possible to Ref.3 and 4
to see an explicit connection with our previous studies.\cite{nis07a}

A simplified linearized set of equations is employed\cite{reduced}
for the numerical simulation
which is reduced from the electron-fluid ion-kinetic hybrid 
gyrokinetic model.\cite{yche01a,lin01,wan01,hin01} 
The equations of Ref.4 are normalized by the ion Larmor radius (at the
electron temperature) for the 
length, the ion cyclotron frequency for time, and the electron temperature 
for the electrostatic potential, and the magnetic field strength at 
the magnetic axis, $B_0$.
The set of the equations are the electron continuity equation 
\be
{\partial \delta{n_e } \over \partial t } = - \nabla_\| \delta{u_{\| e}} 
\label{deltane}
\ee
($ \delta{n_e } $ is the fluid electron density
and $\delta u_{\| e}$ is parallel electron velocity),
the inverse of Faraday's law
\be
{\partial A_\| \over \partial t } =  \nabla_\| \left( \Phi_{eff} - \Phi \right)
\label{faraday}
\ee
($ A_\| $ is the vector potential,
$ \Phi  $ is the electrostatic potential,
and $ \Phi_{eff}  $ is the effective potential 
representing the total parallel electric field),
the gyrokinetic Poisson equation\cite{lee87} 
\be
\Phi - \tilde{\Phi} = \bar{ \delta{n_\alpha}} - \delta{n_e}
\label{poisson}
\ee
($\bar{ \delta{n_\alpha}}$ is the gyro-averaged energetic particle density,
$\tilde{\Phi} $ is the second gyrophase averaged electrostatic potential\cite{lee87})
the lowest order adiabatic relation 
\be
\Phi_{eff}^{} = \delta{n_e},
\label{adiabatic}
\ee
and the inverse of Ampere's law
\be
\delta{u_{\| e}} = \beta_e^{-1} \nabla_{\perp}^2 A_\| + \delta{u_{\| \alpha}}.
\label{ampere}
\ee
The parallel velocity of the energetic particles is given by $\delta{u_{\| \alpha}}$.
Here $\beta_e = {\left( c_{s} / v_{A} \right)}^{2} $ where
$c_s$ is the sound velocity and $v_A$ is the Alfven velocity.
All the variables in Eqs.(\ref{deltane})-(\ref{ampere}) are the
normalized ones. The gradient operators $\nabla_\perp$
and $\nabla_\|$ are those in the direction perpendicular and parallel 
to the equilibrium magnetic field. 

By coupling Eqs.(\ref{deltane})-(\ref{ampere}),
the shear Alfven wave dispersion relation in the toroidal geometry, Eq.(2) of
Ref.3 can be obtained. 
Figure 1 shows the shear Alfven wave frequency as 
a function of the radial coordinate $r$ ($a$ is the minor radius),
which is equivalent to Fig.1 of Ref.3. Due to the $1/R$ variation 
of the toroidal magnetic field ($R$ is the major radius), the cylindrical
Alfven continuum (dashed lines) breaks up and the frequency gap
(or the frequency forbidden band) appears within the range of
$0.299 < \omega / \omega_A < 0.389$. Here, $\omega_A$ is the
Alfven frequency at the magnetic axis. 

Equations (\ref{deltane})-(\ref{ampere}) are employed 
(with the $\bar{\delta{n_{\alpha}}}$ and $\delta{u_{\| \alpha}}$
terms {\it turned off}) in the 
simulation of Fig.~5 of Ref.~4. 
On top of Eqs.(\ref{deltane})-(\ref{ampere}) we add kinetic ions.
The guiding center equation and the $\delta f$ gyrokinetic equation
(the weight equation)\cite{dim92} are 
solved for the kinetic energetic particle ions (we neglect thermal 
kinetic ions, however).
Taking the perturbed distribution function $\delta  f_\alpha $,
the energetic particle density (velocity)
in Eq.\ref{poisson} [Eq.(\ref{ampere})] is given by
$\bar{\delta n_{\alpha}} = r_\alpha \int \delta f_\alpha d^3 v$
($\delta u_{\| \alpha} = r_\alpha \int v_\| \delta f_\alpha d^3 v$),
where $\int d^3 v$ is an average over the velocity space.
Note that in the simulation, we control the perturbed density of
the energetic particles by
multiplying a factor $r_\alpha $ ($r_\alpha < 1$) which is
proportional to the equilibrium energetic particle density.
The energetic ion particles 
are provided with the Maxwellian distribution function
$ f_{0\alpha} \propto \exp{  ( - v_\|^2/ 2 v_\alpha^2)}$ 
in the velocity space 
(thus $\partial f_{0\alpha} / \partial {v_{\|}} $ is always negative).\cite{ros75}
The resonating energetic particles are {\it inserted} within the frequency gap
by utilizing a portion of the Maxwellian distribution function
whose thermal velocity is of the order of Alfven speed.

The particles that resonate with the shear Alfven wave 
with the phase velocity $\omega / k_{\|}$ can destabilize
the TAE mode, when the mode frequency
$\omega$ is within the frequency gap  
(we choose $\omega / \omega_{A} = 0.344 $ in the middle of the gap in Fig.1),
and when the parallel wave vector $k_{\|} = (m - n q ) / q R $ satisfies 
$k_{\|} = - k_{\|m} = k_{\|m+1} $ at $ q = ( 2 m + 1 ) / 2 n $. 
Here, $m$ ($n $) stands for the poloidal (toroidal) mode number.
We take $m=1$, $m+1=2$, and $n=1$ which is equivalent 
to $m=-2$, $m+1=-1$, and $n=-1$ of Ref.3.
The geometrical parameters used for the simulation are the same as in 
Refs.3 and 4 (for example, the inverse aspect ratio of $0.375$
and a parabolic safety factor $q$). 
The major radius is given by $R=46.6 cm$ as well 
(after convincing the TAE excitation in the originally published setting,\cite{nis07a} 
we move on to a parameter survey in a larger size plasma).
From the $\omega $ and the $ k_{\|}$ values chosen above, 
we provide the Maxwellian distribution with 
$v_\alpha = \omega / k_\| = 10.32 c_s$.
The mass and the charge of the energetic particles 
are that of the Hydrogen ion. In the specific simulation below,
we set $\beta_e = 0.01$, and the constant density gradient 
$\kappa_n = -R (1/n_\alpha) (dn_\alpha /dr) = 8.0$.
Here, $n_\alpha$ represents the equilibrium density of the
energetic particles.
The temperature gradient parameters\cite{nis07a} are set to be zero.
In Eq.(\ref{poisson}) and Eq.(\ref{ampere}), $r_\alpha = 0.15$ is taken 
for Figs.2 and 3.

The simulation is conducted by an electromagnetic 
extension\cite{nis07a} of the GTC code\cite{lin98,whi84,lin95}
with a non-iterative field solver.\cite{nis06a,nis06b}
With the additional energetic particle drive, the TAE mode is excited.
A linear eigenmode (contour plot) of the TAE instability is shown in Fig.2.
Note that the contour plots are not up-down symmetric 
due to the finite poloidal shear flow.\cite{kim96}

The frequency spectrum of the TAE instability is shown in Fig.3.
The TAE frequency ($\omega / \omega_A =  -0.36$) is found within
the gap (and not on the gap boundaries as in Ref.4)\cite{psaw}
which is a clear evidence of the TAE excitation.
The linear growth rate of the TAE instability is given
by $\gamma / \omega_A = 0.0215$ (and thus $| \gamma / \omega | = 6.0\%$)
for both the $m=1$ and $m=2$ mode.

Figure 4 shows the linear TAE growth rates (divided 
by the real frequency) versus the multiplication
factor $r_\alpha$. Compared to the calculations
in Figs.2 and 3, a twice larger plasma size is taken for Fig.4  
(the Larmor radius of the energetic particles in Figs.2 and 3
are $15\%$ of the minor radius while $7.5\%$ in Fig.4).
We see a monotonic
increase in the growth rate as the energetic particle 
population increase (and thus the effective beta value of the
energetic particles, $\beta_\alpha$ increases; at $\beta_e = 0.01$,
a simple estimate will give $\beta_\alpha = 4 \pi n_{\alpha} T_\alpha / B_0^2 \sim r_\alpha $.
Here $T_\alpha$ is the energetic particle temperature).
On the other hand, the real frequency of the mode decreases
(approximately $15\%$ of a reduction in the real frequency)
as $r_\alpha$ (or $\beta_\alpha$) increases and crosses
the lower gap boundary\cite{gor98} (but not the upper gap boundary)
which is suggested by the
analysis in Ref.24 (Resonant TAE, which emphasizes the resonance
between the mode frequency and the magnetic drift frequency of the 
energetic particles).\cite{che95} 
As a reference, the energetic particle mode (EPM)\cite{zon96}
refers to a heating of the continuum based on the notion that
the energetic particle drive exceeds the continuum damping
and predicts the appearance of the mode
frequency both in the upper and the lower continuum.
The square plots in Fig.4 represents the 
analysis of the TAE growth rate 
in a large aspect ratio tokamak from Ref.3 
(the simulation results and the analysis compare favorably 
at higher $\beta_\alpha$).
 
We note that instability growth was already minimal at $r_\alpha = 0.025$
(with the specific simulation parameters we employed in Fig.4),
and we did not survey below $r_\alpha < 0.025$ in this work. 
The TAE mode in its nature should not have
an instability threshold.
The latter onset feature needs to be investigated in detail
to see the limitation of the initial value approach (if any).
We also would like to remind that a simplified model Eqs.(1)-(5) is employed
in this letter (so as to primary focus on the excitation of TAE by
the additional energetic particles).\cite{nis07a}
The radial extension of the simulation domain is limited to $ 0.1 < r/a < 0.9$ (see Fig.2).
An inclusion of the magnetic axis can be crucial to
describe the long wavelength global modes precisely.

In summary, the first linear excitation of the low-n 
TAE modes by the energetic particles 
in a global gyrokinetic particle simulation is reported.\cite{mis08}
The work did not employ MHD model (through closure relations).
With a completion of the current
global gyrokinetic simulation method, one can investigate
the onset and the saturation mechanism of the TAE modes simultaneously 
without any restrictions on the wavelength of the modes.
Apparently, the advantage of initial value approach
is its application for nonlinear simulation.
We plan to report
the analysis of energetic particles driven high-n Alfvenic modes separately.
Whether which mode numbers are most unstable is a great
interest to large tokamak burning plasma experiments.

The author would like to
thank Dr.~Z.~Lin, Dr.~W.~X.~Wang, Dr.P.~H.~Diamond, Dr.~T.~S.~Hahm, Dr. B.~Scott,
Dr. M.~Yagi, Dr.K.~C.~Shaing, and Dr. C.~Z.~Cheng for discussions.
This work is supported by Department of Energy (DOE) SciDAC 
Center for Gyrokinetic Particle Simulation and National Cheng Kung University
Top University Project.
The simulation is done employing National Energy Research Scientific
Computing Center (NERSC) supercomputers by the year 2007
during YN's residence at the University of California, Irvine.

\begin{figure}
    \includegraphics[width=7.cm,angle=+0] {dummy.eps}
    \includegraphics[width=9.cm,angle=+0] {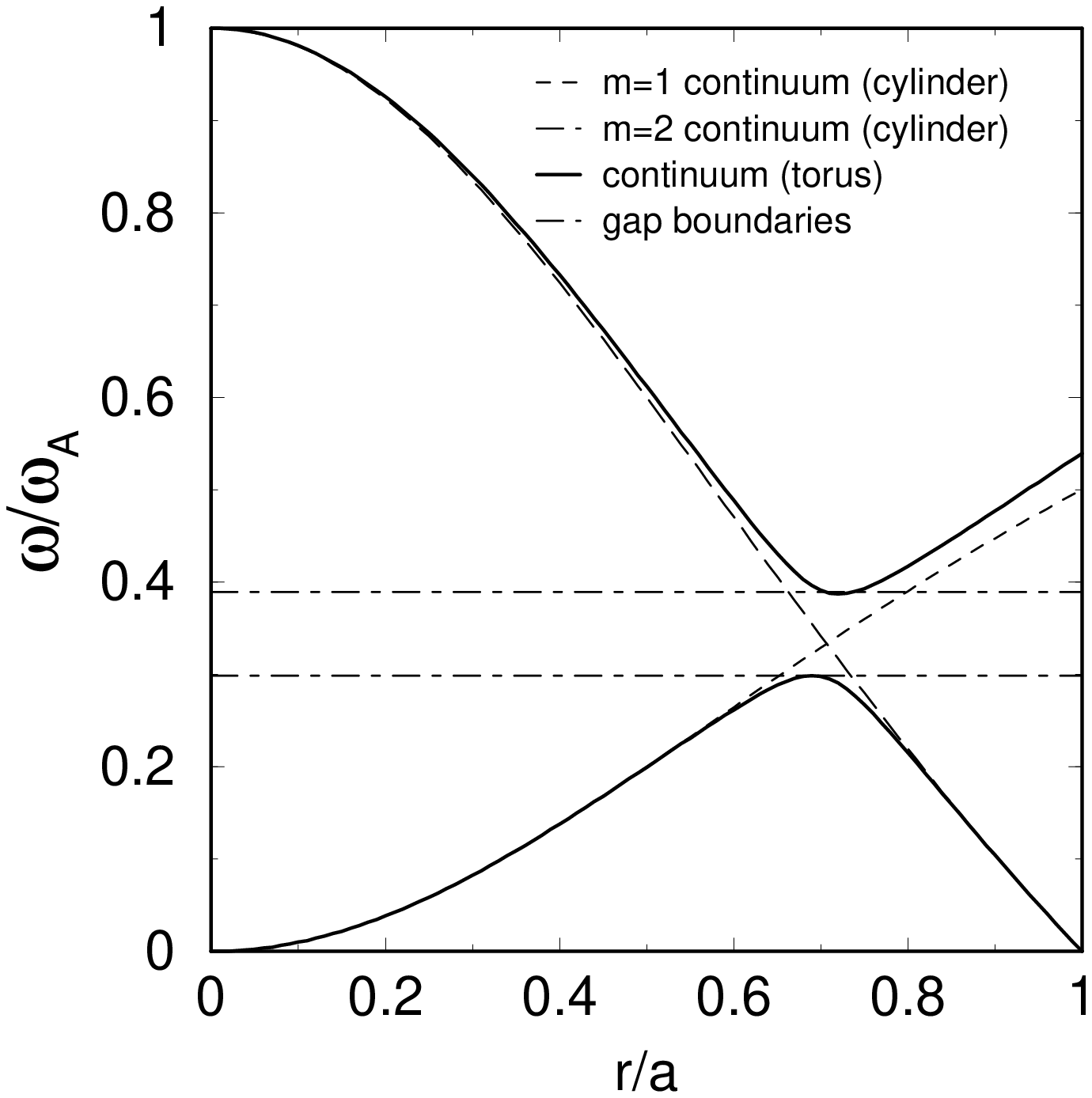}
\caption{Shear Alfven frequency as a function of the radial location.
The dashed curves are the continuum frequencies for the cylindrical limit,
for $m=1$ and $m=2$ modes.
The solid lines are for the continuum
frequency with the toroidal geometry effect.
The lower (upper) boundary of
the upper (lower) curve is at $\omega/ \omega_{A} = 0.389$ ($\omega/ \omega_{A} = 0.299$).
Correspondingly the frequency gap (the forbidden frequency range)
appears within the range of $0.299 < \omega/ \omega_{A} < 0.389$.
The figure is a recapitulation of Fig.1 of Ref.3.  }
\label{fig1}
\end{figure}

\newpage
\vspace{-2cm}
\begin{figure}
    \includegraphics[width=4.cm,angle=+0] {dummy.eps}
    \includegraphics[width=4.cm,angle=+0] {dummy.eps}
    \includegraphics[width=8.cm,angle=+0] {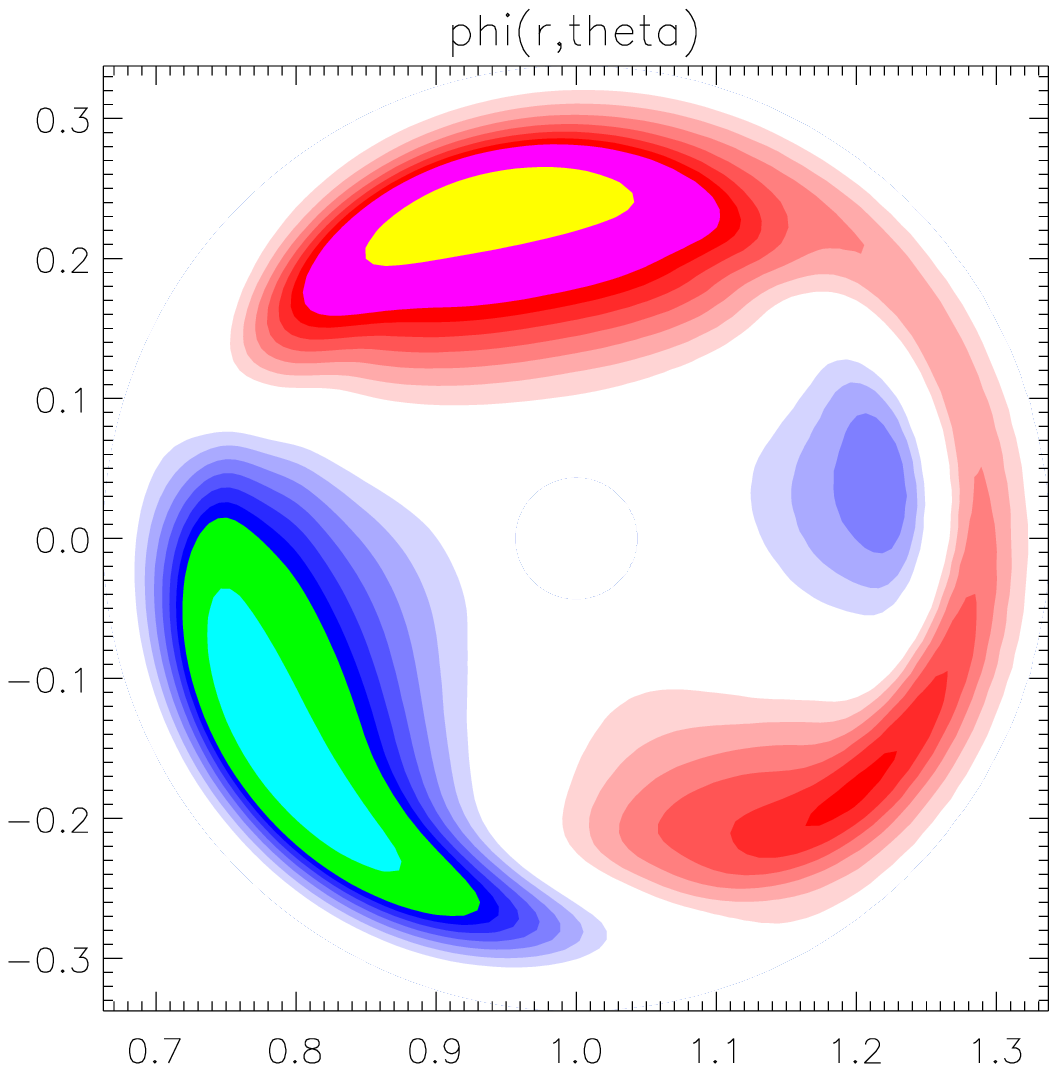}
    \includegraphics[width=8.cm,angle=+0] {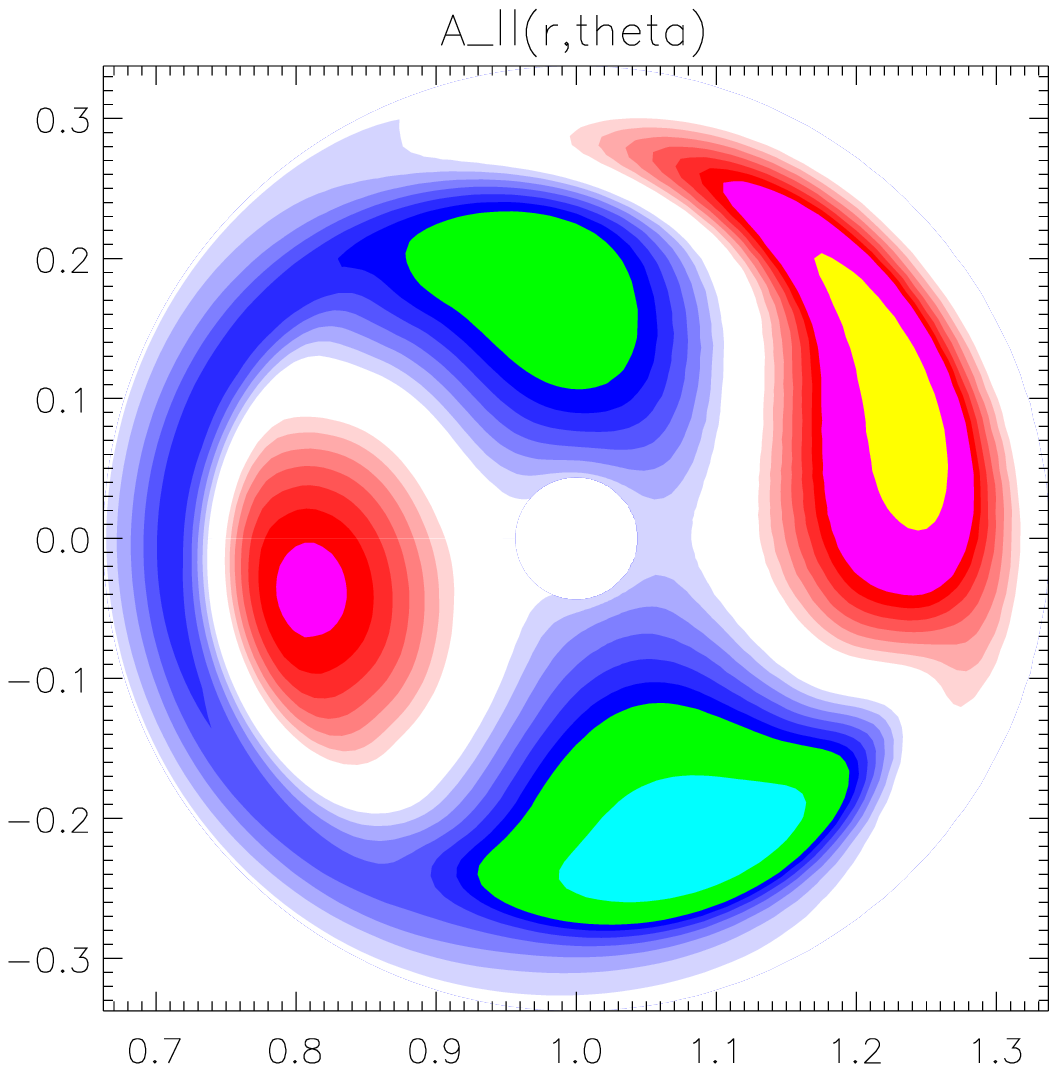}
\caption{ (Color online)
Linear eigenmodes (contour plots on a poloidal plane) of TAE instability.
(a) The electrostatic potential $\Phi$ 
at a toroidal angle $\zeta=0$.
Red represents positive $\Phi$ values, while blue represents
negative $\Phi$ values.
(b) The vector potential $A_\|$ at a toroidal angle $\zeta=0$.}
\label{fig2}
\end{figure}

\newpage
\begin{figure}
    \includegraphics[width=4.cm,angle=+0] {dummy.eps}
    \includegraphics[width=4.cm,angle=+0] {dummy.eps}
    \includegraphics[width=9.cm,angle=+0] {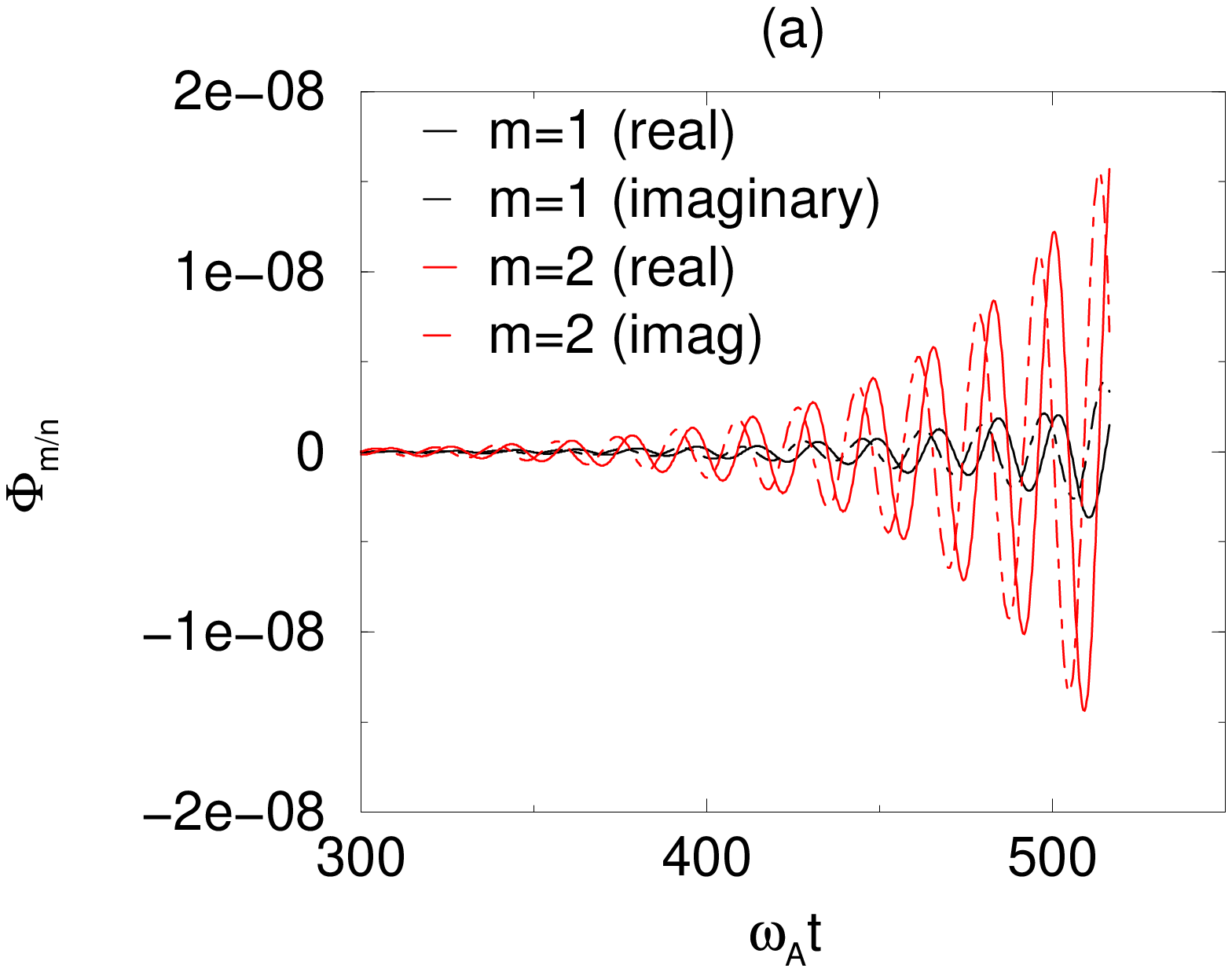}
    \includegraphics[width=8.cm,angle=+0] {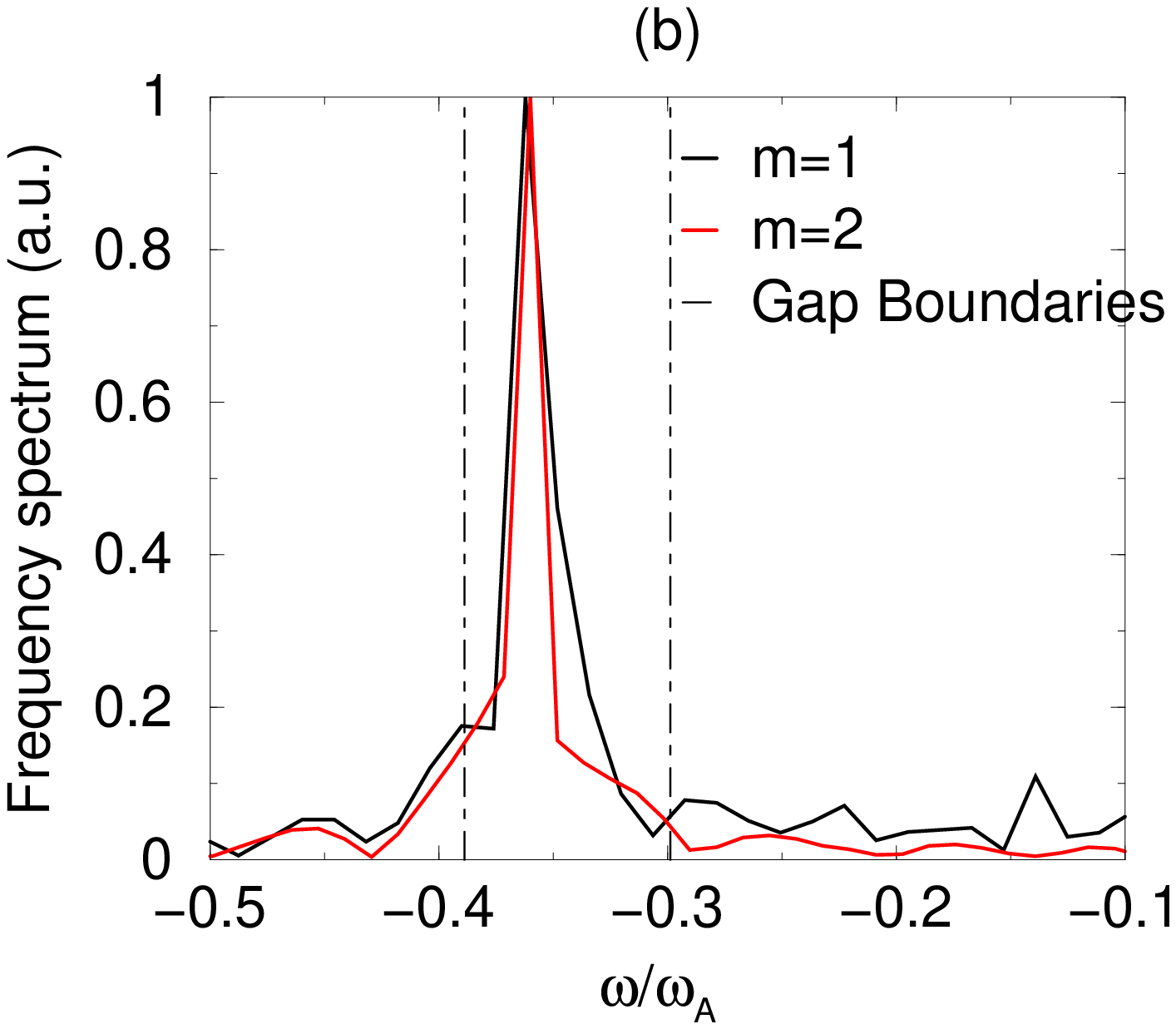}
\caption{ (Color online)
(a) The Fourier components of the electrostatic potential 
$\Phi$ as a function of time.
(b) The frequency spectrum obtained from the time series
of Fig.3(a) (black for $m=1$, red for $m=2$).}
\label{fig3}
\end{figure}

\begin{figure}
    \includegraphics[width=7.cm,angle=+0] {dummy.eps}
    \includegraphics[width=9.cm,angle=+0] {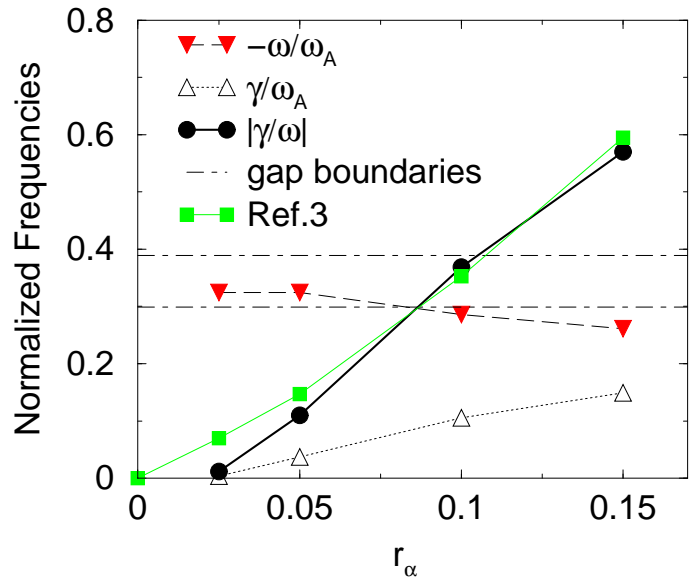}
\caption{ (Color online)
Dependence of the real frequency $\omega$, and the 
linear growth rate $\gamma$ on $r_\alpha$ (and thus $\beta_a$).
For each data, eigenmode structures similar to Fig.2 are obtained.
The green squares represent the analytical TAE growth rate -
real frequency ratio in a large aspect ratio tokamak from Ref.3.}
\label{fig4}
\end{figure}


\begin{thebibliography}{99}
\bibitem{reduced} These simplified equations are employed for the simulation of the 
Alfven eigenmodes in Ref.4 (but not for the micro-instability part).
\bibitem{psaw} We remind that the upper branch of the continuum
accumulates more prominently than the lower branch
as predicted at the zero resistivity limit
of the periodic shear Alfven wave mode.\cite{che85,che95} See Fig.5(b) of Ref.4.
\end{thebibliography}
\end{document}